\documentclass[lettersize,journal]{IEEEtran}
\usepackage{amsmath,amsfonts}
\usepackage{algorithmic}
\usepackage{array}
\usepackage{textcomp}
\usepackage{stfloats}
\usepackage{url}
\usepackage{upgreek}
\usepackage{verbatim}
\usepackage{svg}
\usepackage{graphicx}
\usepackage{tikz}
\usepackage{pifont}
\usepackage[caption=false,font=footnotesize]{subfig} 

\DeclareUnicodeCharacter{2212}{$-$}

\def\BibTeX{{\rm B\kern-.05em{\sc i\kern-.025em b}\kern-.08em
    T\kern-.1667em\lower.7ex\hbox{E}\kern-.125emX}}
\usepackage{booktabs}
\usepackage{makecell}
\usepackage{multirow}
\usepackage{balance}
\begin{document}
\title{OwnDPDLab: A Flexible Open-Source Testbed for Wideband 
DPD Algorithm Benchmarking}
\author{Marvin Jaeger,~\IEEEmembership{Graduate Student Member,~IEEE}, Philipp Luetke, David Kopyto, Georg Frederik Riemschneider,\\ Omar Jabi, and Alexander Koelpin,~\IEEEmembership{Fellow,~IEEE}

\thanks{Received XXXXXX XX, 2026;
This work is part of the Hamburg Quantum Computing (HQC) project, where the required technologies for the implementation of a quantum computer are developed. The project is co-financed by ERDF and Fonds of the Hamburg Ministry of Science, Research, Equalities and Districts (BWFGB). (\textit{Corresponding author: Marvin Jaeger}.)\\
M. Jaeger, G. F. Riemschneider, O. Jabi, and A. Koelpin, are with the Institute of High-Frequency Technology, Hamburg University of Technology, Hamburg, Germany
(e-mail: marvin.jaeger@tuhh.de).\\
P. Luetke is with the Smart Sensors Group, Hamburg University of Technology, Hamburg, Germany and the eleqtron GmbH, Hamburg, Germany.\\
D. Kopyto is with the Institute of Communications Technology, Hamburg University of Technology, Hamburg, Germany.\\
Digital Object Identifier XX.XXXX/XXXX.2026.XXXXXXX}
}

\markboth{ }%
{Jaeger \MakeLowercase{\textit{et al.}}: OwnDPDLab: A Flexible Open-Source Testbed for Wideband DPD Algorithm Benchmarking on RFSoC}

\maketitle
\bstctlcite{IEEEexample:BSTcontrol}
\begin{abstract}

5G and Beyond-5G standards require digital predistortion (DPD) algorithms to operate on increased signal bandwidths. Wideband
laboratory test hardware is cost-intensive, and openly available solutions lack flexibility. The OwnDPDLab provides a highly
flexible, affordable, open-source, and openly accessible system. It is based on the RFSoC\,4x2 and supports full control of center frequency, sampling mode, output power, and input attenuation at a signal bandwidth of up to 1\,GHz. The system's capability is demonstrated by linearizing a laboratory power amplifier using a 196.608\,MHz orthogonal frequency division multiplexing (OFDM) signal with 256-QAM modulation using both a memory polynomial and an augmented real-valued time-delay neural network in the first and second Nyquist zone. The system achieves a normalized mean squared error improvement of up to 23\,dB and an adjacent channel leakage ratio improvement of up to 11\,dB, using DPD. 
\end{abstract}


\section{Introduction}

\IEEEPARstart{D}{igital} predistortion (DPD) is the de facto method in communication engineering to compensate the nonlinear behavior and memory effects of the power amplifier (PA) \cite{PALinJournal}. Additionally, DPD is also shown to be relevant in quantum computing \cite{jaeger2026DPD,keysigth2026DPD}, where it is used to linearize microwave frontends \cite{GEMIC25} using a feedback receiver \cite{GEMIC26_2}. Both fields require increased bandwidth and control precision and therefore increased requirements for the DPD algorithms.

The algorithms can be separated in polynomial based equations like the general memory polynomial (GMP) \cite{DPD_GMP} or its reduced variant the memory polynomial (MP) \cite{DPD_MP}. Latest research deals with neural network based DPD, considering feedforward networks \cite{DPD_ANN}, residual neural networks \cite{DPD_AN_Residual}, and feedback based neural networks \cite{DPD_LSTM}.

While commercial laboratory equipment and software defined radios with up to 1\,GHz bandwidth are costly, low-cost equipment lacks bandwidth and signal quality or reliability on the data transmission. The online available and open to use RFWebLab \cite{DPD_comp15} is a good starting point for testing DPD algorithms, but lacks flexibility, and availability is not guaranteed. The OpenDPD framework \cite{OpenDPD} is also free to use, and supports real measurement data for offline training but does not allow for online verification. Laboratory systems like \cite{Keysight_M5000_2026,niPXIe5842}, and professional software defined radio (SDR) solutions \cite{usrpX440} support high bandwidth and high channel quality, but are at high costs. The open-source SDR~\cite{siauciulis2023} is built on the RFSoC platform, but its signal bandwidth is limited to 61.44\,MHz due to the network interface bottleneck. This necessitates an affordable and flexible wideband solution especially designed to fulfill the reliability requirements for training DPD algorithms. Table~\ref{tab:comparison} summarizes this comparison.

This paper presents the OwnDPDLab, an openly available and inexpensive alternative based on a radio frequency system-on-chip (RFSoC)\,4x2 evaluation board \cite{RFSoC4x2}. This system is especially developed for DPD testing and an overview is depicted in Fig.~\ref{fig:SystemPlan}. However, it can also be used for other applications in communication and microwave engineering. The main contributions of this paper are as follows: we present an open-source, low-cost DPD testbed based on the RFSoC\,4x2 evaluation board, providing a signal bandwidth of 983.04\,MHz at an affordable cost accessible to standard academic laboratories; we design the system to provide high flexibility and precise data synchronization, enabling control of the center frequency, output power, and input attenuation; and we provide an experimental comparison of a memory polynomial (MP) and an augmented real-valued time-delay neural network (ARVTDNN) in both the first and second Nyquist zone of the RFSoC, demonstrating the testbed's capability to benchmark state-of-the-art DPD algorithms. 

\begin{table}[tb]
	\centering
	\scriptsize
	\caption{Comparison of Wideband Test Systems}
	\label{tab:comparison}
	\begin{tabular}{@{}lcccc@{}}
		\toprule
		\textbf{System} & \textbf{Inst. BW} & \textbf{Fc/VOP} & \textbf{Cost} & \textbf{Open source} \\
		\midrule
		RFWebLab \cite{DPD_comp15}       & 200\,MHz     & \ding{55} & free use  & \ding{55} \\
		OpenDPD \cite{OpenDPD}           & -- (offline)   & \ding{55} & free   & \ding{51} \\
		RFSoC-GNURadio \cite{siauciulis2023} & 61.44\,MHz     & \ding{51} & low  & \ding{51} \\
		Ettus USRP X440 \cite{usrpX440}        & 1.6\,GHz          & \ding{51} & high & \ding{51} \\
		Lab Systems \cite{Keysight_M5000_2026,niPXIe5842} & $>$1\,GHz & \ding{51} & high & \ding{55} \\
		\midrule
		\textbf{This Work}        & \textbf{983.04\,MHz} & \ding{51} & \textbf{low} & \ding{51} \\
		\bottomrule
	\end{tabular}
\end{table}

\section{Testbed Architecture}

\begin{figure*}
	\centering
	\footnotesize
	\def\svgwidth{\textwidth}
	\includesvg{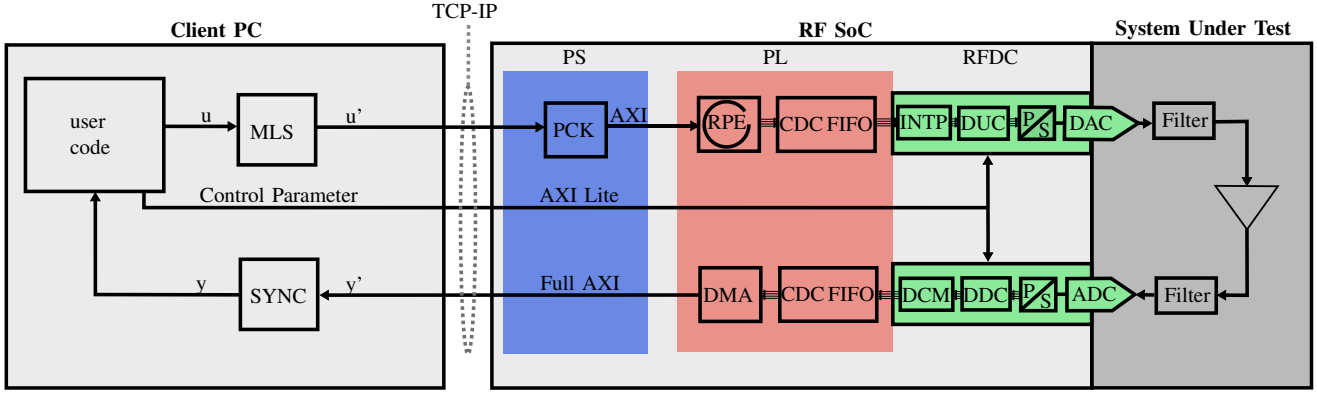}
	\caption{Detailed system overview of the DPD testbed based on an RF SoC and depicting the data flow from the user code to the RFSoC to the amplifier and back, as well as the internal processing and signal flow within the RFSoC.}
	\label{fig:SystemPlan}
\end{figure*}

Fig. \ref{fig:SystemPlan} shows a system overview of the DPD testbed. Its parameters can be controlled from the connected PC. The user's test in-phase quadrature (IQ) samples, offline calculated on a personal computer (PC), are sent to the RFSoC via transmission control protocol/internet protocol (TCP-IP). The samples are received on the processing system (PS) of the RFSoC, packaged (PCK), and transmitted via advanced extensible interface (AXI) interconnect to the programmable logic (PL). An ultra RAM (URAM)-based replay engine receives the data from the PS and replays them cyclically in a parallel AXI stream using a super sampling rate of SSR\,=\,4. The parallel structure allows the system to increase the output signal bandwidth, without increasing the system clock. A clock domain crossing first in, first out (CDCFIFO) buffer ensures a sample-loss-free clock domain crossing. The supersampled baseband signal is transferred to the output radio frequency data converter (RFDC), which consists of an interpolator (INTP), a digital upconverter (DUC), a serializer (P/S), and the digital-to-analog converter (DAC). Additionally, the output power of the included DAC is controllable. It outputs the sampled signal, and an external analog filter selects the used Nyquist zone. The amplified signal is observed by the RFDC receive channel, which consists of an analog-to-digital converter (ADC) with controllable attenuation, a deserializer (S/P), a digital downconverter (DDC), and a decimation filter (DCM). The input RFDC outputs the data to the PL with the super sampling rate of SSR\,=\,4. This signal is transferred via a direct memory access (DMA) controller to the PS and received by the PC via TCP-IP.

This setup is based on the RFSoC\,4x2 \cite{RFSoC4x2}, configured with a sampling rate $R$ of 4.9152\,GSPS for the ADC and the DAC. The internal clock is set to 245.76\,MSPS with SSR\,=\,4, yielding an effective signal bandwidth of 983.04\,MHz. The replay engine (RPE) is designed to store 880\,000 IQ samples.

The DUC enables free up- and downconversion within the first Nyquist zone from 0 to 2\,457.6\,MHz. Higher center frequencies are accessible by selecting a higher Nyquist zone via an external bandpass filter, eliminating the need for an external up-converter. The DAC is configured in mixed mode for this purpose. Due to the sinc-shaped frequency response of the DAC, signal attenuation increases with higher Nyquist zones. Additionally, the sinc-shaped attenuation distorts the in-band signal. For compensation, the hardcoded inverse sinc filter of the RFSoC is activated, which automatically adapts to the selected sampling mode and pre-equalizes the signal for both Nyquist zones.

Adjusting the DAC's variable output power (VOP) shifts the operating point of the PA without reducing the digital signal amplitude, which would otherwise decrease the signal-to-quantization-noise ratio (SQNR). The ADC's digital stepped attenuator (DSA) can also be adjusted by the user to prevent saturation and reduce the need for external attenuators.

The synchronization is done on the client by prepending an 
adjustable maximum length sequence (MLS) to the excitation 
signal. Both the observation signal $y'[n]$ and the reference 
MLS are upsampled by factor $I$, yielding $y_{\mathrm{int}}[k]$ 
and $\text{MLS}_{\mathrm{int}}'[k]$. The sub-sample delay 
index $\hat{k}$ is then extracted via cyclic cross-correlation 
in the frequency domain:
\begin{equation}
    \hat{k} = \underset{k}{\mathrm{arg\,max}} \, \mathcal{F}^{-1} 
    \left\{ \mathcal{F}\{\text{MLS}_\mathrm{int}'\}^* \odot 
    \mathcal{F}\{y_{\mathrm{int}}\} \right\}
\end{equation}
where $\mathcal{F}$, $(\cdot)^*$, and $\odot$ denote the FFT 
operator, complex conjugate, and element-wise multiplication, 
respectively.

\section{Experimental Demonstration}

This section demonstrates the abilities of the presented testbed and gives an exemplary comparison of a polynomial based DPD and a neural network based DPD in the first and second Nyquist zones.

\subsection{Measurement Setup}

The testbed linearizes a laboratory wideband PA (Mini-Circuits ZHL-42-SMA, 0.7--4.2\,GHz) operating at 900\,MHz in the first and 3\,915.2\,MHz in the second Nyquist zone of the RFSoC. To select the desired Nyquist zone and suppress aliasing, the signal path is filtered with a lowpass filter (Mini-Circuits VLFX-2500+, $f_c = 2.5$\,GHz) in the first zone and a bandpass filter (Mini-Circuits VBFZ-3590-S+, 3--4.3\,GHz) in the second zone. Coaxial attenuators providing 17\,dB of attenuation protect the RFSoC input from the amplified PA output.

The DAC is set up to an average output power of 2\,dBm for the 1st Nyquist zone and 4.5\,dBm for the 2nd Nyquist zone to compensate for the sinc roll-off. The DSA is set to an input attenuation of 8\,dB. The DUC and DDC frequencies are set to 0.9\,GHz for the first Nyquist zone and -1\,GHz for the second Nyquist zone, respectively. The negative sign prevents a mirrored spectrum. The DAC mode is set to sample and hold for the first Nyquist zone and in mixed mode for the second Nyquist zone.

The test signal is a 256-QAM OFDM waveform with a bandwidth of 
196.608\,MHz, which constitutes exactly one-fifth of the total 
sampling rate of 983.04\,MSPS. Therefore, its fifth-order 
intermodulation (IM5) products expand to exactly fit the available 
observation bandwidth without inducing any out-of-band aliasing 
artifacts. We use 200\,000 samples for live testing
on the amplifier.

\begin{table}[tb]
\centering
\caption{Summary of Optimized DPD Parameters}
\begin{tabular}{@{}llrr@{}}
\toprule
\textbf{Model} & \textbf{Parameter} & \textbf{Nyquist Zone 1} & \textbf{Nyquist Zone 2} \\ \midrule
\multirow{2}{*}{\textbf{MP}}
 & $K, L$ & 5, 9 & 3, 8 \\
 & Complexity (Coeffs.) & 30 & 18 \\ \midrule
\multirow{4}{*}{\textbf{ARVTDNN}}
 & $K', L'$ & 5, 9 & 3, 8 \\
 & Hidden Sizes & [128, 64] & [64] \\
 & Activation & SiLU & SiLU \\
 & Complexity (Params) & 17\,474 & 3\,074 \\ \bottomrule
\end{tabular}
\label{tab:DPD_coef}
\end{table}

\subsection{Exemplary DPD Performance Comparison}
The input and received output samples are compared and an inverse relation is estimated. In the first case, a MP \cite{DPD_MP} is applied. It models the nonlinearity and memory effects. It is defined as
\begin{equation}
	u[n] = \sum_{\substack{k=1 \\ k \text{ odd}}}^{K} 
             \sum_{l=0}^{L} h_{k,l}\, x[n-l]\, |x[n-l]|^{k-1} \label{eq:mp}
\end{equation}
where $K$ is the nonlinearity order, $L$ is the memory depth, $h_{k,l}$ are the coefficients of the model, and $x[n]$ is the input signal. The coefficients are estimated using the least squares (LS) method.

To compare this polynomial based method with a neural network based method, the ARVTDNN is used \cite{DPD_ANN}. This network uses a tapped delay line (TDL) structure with depth $L'$, computing for each tap the in-phase component, quadrature component, and magnitude power terms $|x|^1$ through $|x|^{K'}$. Additional nonlinearity and cross-terms are introduced by the fully connected layers and the activation function.

To find the best hyperparameters, we use a parameter search based on tree-structured parzen estimator \cite{Optuna}, optimizing for the ACLR value, ensuring the NMSE is below -35\,dB. The optimizer uses 100\,000 recorded input and output samples. It uses up to 2000 iterations for the MP and 5000 iterations for the ANN, respectively. Table~\ref{tab:DPD_coef} presents the chosen parameters within the search space.

The experiment is executed using those hyperparameters. The amplifier is measured without DPD and normalized by the LS estimation of the complex gain $G$. Both models are trained using 100\,000 normalized samples. After training, the models are validated by predistorting the full dataset, applying hard clipping to prevent DAC saturation, and applying them to the system. The experimental code is available at \cite{JaegerOwnDPDLab}.

{{\begin{figure}
	\centering
	\footnotesize
	\def\svgwidth{0.9\columnwidth}
	\includesvg{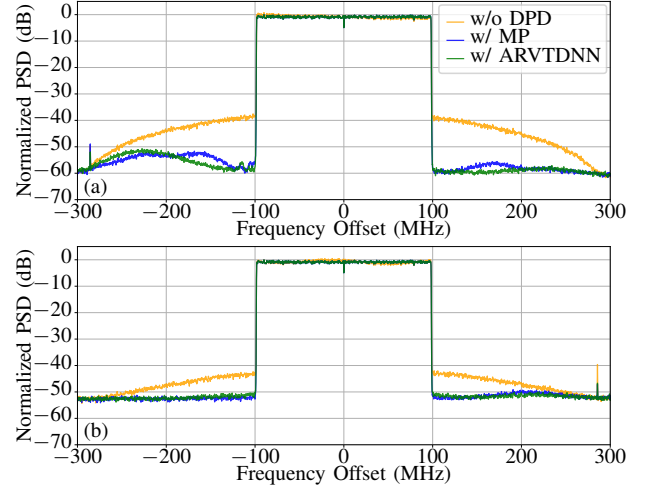}
	\caption{Spectra of the test signal without DPD, with MP, and ARVTDNN in the (a) first and (b) second Nyquist zone.}

	\label{fig:spectrum}
\end{figure}}
}

Fig.~\ref{fig:spectrum} shows the spectra of the test signal without DPD, with MP, and ARVTDNN in the first and second Nyquist zone. It demonstrates that both DPD algorithms are able to linearize the PA and therefore reduce the out of band emissions, as well as the inband distortion. In Fig.~\ref{fig:spectrum}(a) the ARVTDNN shows a stronger out of band suppression for positive offset frequencies, but lower suppression for negative offset frequencies than the MP. Fig.~\ref{fig:spectrum}(b) shows the spectra in the second Nyquist zone. The PA operates at a reduced operation point, due to the sinc roll-off as well as higher transmission losses caused by higher frequencies. This leads to reduced nonlinear effects of the PA and diminishes differences between the ARVTDNN and the MP in compensating the PA nonlinearities. However, due to reduced power, the SNR is reduced, leading to a higher noise floor in the spectrum, limiting out-of-band reduction performance. The discrete spurious tone observed in the spectrum originates from clock feedthrough and parasitic coupling of internal sub-harmonic clock dividers into the analog RF output path.

{{\begin{figure}
	\centering
	\footnotesize
	\def\svgwidth{1\columnwidth}
	\includesvg{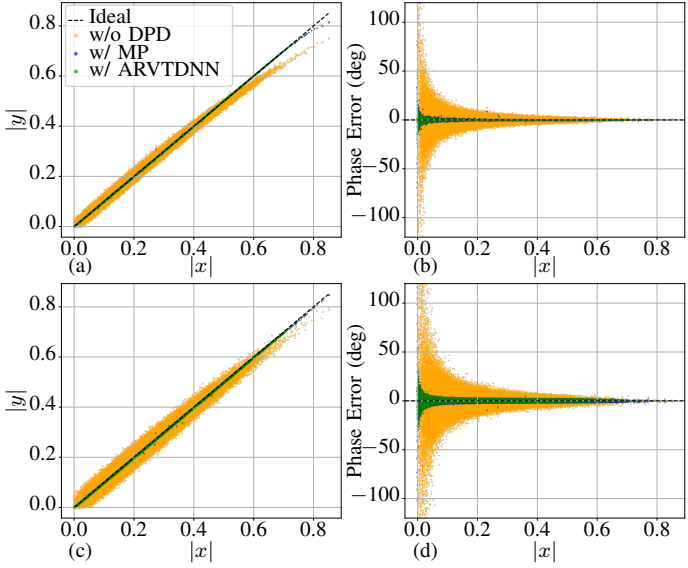}
	\caption{AM-AM and AM-PM characteristics of the test signal without DPD, with MP, and ARVTDNN in the (a) and (b) first and (c) and (d) second Nyquist zone, respectively.}
	\label{fig:AMAM_AMPM}
	\vspace{-5mm}
\end{figure}}}

Fig.~\ref{fig:AMAM_AMPM}(a) and (b) show the AM-AM and AM-PM 
in the first Nyquist zone. Without DPD, both characteristics 
show memory-induced spread around the ideal line, as well as 
compression at high amplitudes. Both DPD methods reduce this 
spread similarly, though a residual saturation persists at 
maximum amplitude, as PA compression cannot be fully inverted.
Fig.~\ref{fig:AMAM_AMPM}(c) and (d) show the second Nyquist 
zone. It shows a wider phase distribution across all normalized amplitudes. This 
results from the lower absolute operating power, which shifts 
the PA operation into a different region of the phase-amplitude 
characteristic when normalized to the signal's peak amplitude.

To quantify the results in metrics, Table~\ref{tab:DPD_results} summarizes the normalized mean squared error (NMSE), adjacent channel leakage ratio (ACLR), and error vector magnitude (EVM). It shows a comparable improvement using a MP and the ARVTDNN, where the network requires more resources for implementation and is more computationally complex. However, both methods achieve comparable linearization performance, while the MP requires significantly fewer parameters, demonstrating that polynomial methods remain competitive at moderate PA nonlinearity levels. 
Additionally, the improvement achieved by DPD is more pronounced in the first Nyquist zone, consistent with the higher operating point.

\begin{table}[h]
\centering
\caption{DPD Performance Comparison Across Nyquist Zones}
\label{tab:DPD_results}
\begin{tabular}{lll}
\toprule
\textbf{Method} & \textbf{Nyquist Zone 1} & \textbf{Nyquist Zone 2} \\
\midrule
w/o DPD  & \makecell[l]{NMSE: $-$24.47\,dB \\ ACLR: $-$41.75\,dBc \\ EVM:\phantom{0}5.98\,\%}
         & \makecell[l]{NMSE: $-$19.95\,dB \\ ACLR: $-$44.97\,dBc \\ EVM: 10.06\,\%} \\
\addlinespace
MP       & \makecell[l]{NMSE: $-$47.29\,dB \\ ACLR: $-$52.52\,dBc \\ EVM:\phantom{0}0.43\,\%}
         & \makecell[l]{NMSE: $-$35.38\,dB \\ ACLR: $-$49.33\,dBc \\ EVM:\phantom{0}1.70\,\%} \\
\addlinespace
ARVTDNN  & \makecell[l]{NMSE: $-$46.28\,dB \\ ACLR: $-$52.47\,dBc \\ EVM:\phantom{0}0.49\,\%}
         & \makecell[l]{NMSE: $-$35.80\,dB \\ ACLR: $-$49.53\,dBc \\ EVM:\phantom{0}1.62\,\%} \\
\bottomrule
\end{tabular}
\end{table}

\section{Conclusion}\label{sec:Conclusion}
This paper describes the need for a low-cost DPD testbed and proposes the OwnDPDLab. The system is based on the RFSoC 4x2 and is therefore affordable and makes DPD algorithms for communication engineering or quantum computing verifiable and reproducible. This system also contains a transceiver frontend which can be controlled by the user. Additionally, higher Nyquist zones are available for use. The system allows for the generation of 1\,GHz signal bandwidth. This work demonstrates its abilities on a 196.608\,MHz OFDM signal using 256 QAM. It was demonstrated to linearize a PA for this signal in the first and second Nyquist zone comparing a MP and an ARVTDNN. The MP shows a comparable performance with lower complexity compared to the ARVTDNN.  The testbed's client and PS code are open sourced, as well as the PL bitstream, and available on GitHub \cite{JaegerOwnDPDLab}. Future work may include pursuing a theoretical bandwidth of 2.5\,GHz by increasing the super sampling factor and system clock. Additionally the number of available DAC and ADC channels will be increased, to enable multichannel DPD testing. The framework will be extended to additional evaluation boards from Xilinx.

\section*{Acknowledgment}
The authors used Claude (Anthropic) to assist with editing the manuscript text and with parts of the testbed code. All content was reviewed and verified by the authors, who remain fully responsible for it.
\bibliographystyle{IEEEtran}

\bstctlcite{IEEEexample:BSTcontrol}
\bibliography{MyBib_clean,BSTcontrol.bib}

\end{document}